\PassOptionsToPackage{colorlinks=true,linkcolor=blue,citecolor=blue,urlcolor=blue}{hyperref}
\documentclass[sigconf,nonacm]{acmart}

\usepackage{pgfplots}
\pgfplotsset{compat=1.18}
\usepackage{caption}
\usepackage{subcaption}
\usepackage{subcaption}
\usepackage{tikz}
\usepackage{svg}
\usepackage{colortbl}  
\usepackage{booktabs}   
\usepackage[table]{xcolor}
\usepackage{multirow}
\usepackage{tabularx}
\usepackage{array}
\usepackage{tcolorbox}
\usepackage{listings}
\renewcommand{\arraystretch}{1.45}
\usetikzlibrary{patterns.meta}

\usetikzlibrary{patterns}

\title{Game-Master LLMs for Task-Based Role-Play: Supporting the Acquisition of Idiomatic Language in L2 Learning}

\author{Amir Tahmasbi}
\affiliation{
  \institution{Purdue University}
  \city{}
  \country{}
}
\email{atahmasb@purdue.edu}
\author{Milad Esrafilian}
\affiliation{
  \institution{Purdue University}
  \city{}
  \country{}
}
\email{mesrafil@purdue.edu}
\author{Judson Wright }
\affiliation{
  \institution{Purdue University}
  \city{}
  \country{}
}
\email{wrigh703@purdue.edu}
\author{Sooyeon Jeong }
\affiliation{
  \institution{Purdue University}
  \city{}
  \country{}
}
\email{sooyeonj@purdue.edu}
\author{Aniket Bera }
\affiliation{
  \institution{Purdue University}
  \city{}
  \country{}
}
\email{aniketbera@purdue.edu}

\begin{document}
\hypersetup{
    colorlinks=true,
    linkcolor=blue,
    citecolor=blue,
    urlcolor=blue
}

\begin{abstract}

Natural and idiomatic expressions are essential for fluent, everyday communication, yet many second-language learners struggle to acquire and spontaneously use casual slang despite strong formal proficiency. To address this gap, we designed and evaluated an LLM-powered, task-based role-playing game\footnote{Our game is available at \url{https://ashenrealms.me/}} in which a GPT-4o-based Game Master guides learners through an immersive, three-phase spoken narrative. After selecting five unfamiliar slang phrases to practice, participants engage in open-ended dialogue with non-player characters; the Game Master naturally incorporates the target phrases in rich semantic contexts (implicit input enhancement) while a dedicated Practice Box provides real-time explicit tracking and encouragement. Post-session, learners receive multi-level formative feedback analyzing the entire interaction.

We evaluated the system in a between-subjects study with 14 international graduate students, randomly assigned to either the RPG condition or a control condition consisting of a traditional AI-led virtual classroom. Results from an immediate post-test show that the RPG group achieved greater gains in both comprehension of the target phrases and their accurate, contextual use in sentences. A one-week delayed post-test further demonstrates that these gains are retained over time, with the RPG group showing a 21--27\% improvement, indicating the effectiveness of our approach in supporting longer-term learning. Qualitative survey responses assessing engagement and perceived effectiveness further indicate that the game-based approach provided more practice opportunities and a more natural learning experience. These findings highlight the potential of narrative-driven LLM interactions in vocabulary acquisition.

\end{abstract}

\maketitle
\section{Introduction}
Language is a fundamental tool for human communication, enabling individuals to share information, convey ideas, and interact socially and professionally \cite{shashkevich2019power,Renau2016ARO}. In a globalized world, effective communication across cultural and linguistic boundaries is essential for both collaboration and coexistence. Universities provide a clear illustration of this need, as individuals from diverse backgrounds come together to learn, research, and exchange ideas. In the 2023-2024 academic year alone, over 1.1 million international students were enrolled in U.S. higher education institutions, comprising approximately 6\% of the total student population \cite{iie2023opendoors}. 
\begin{figure}[htbp]
    \centering
    \includegraphics[width=1.2\columnwidth]{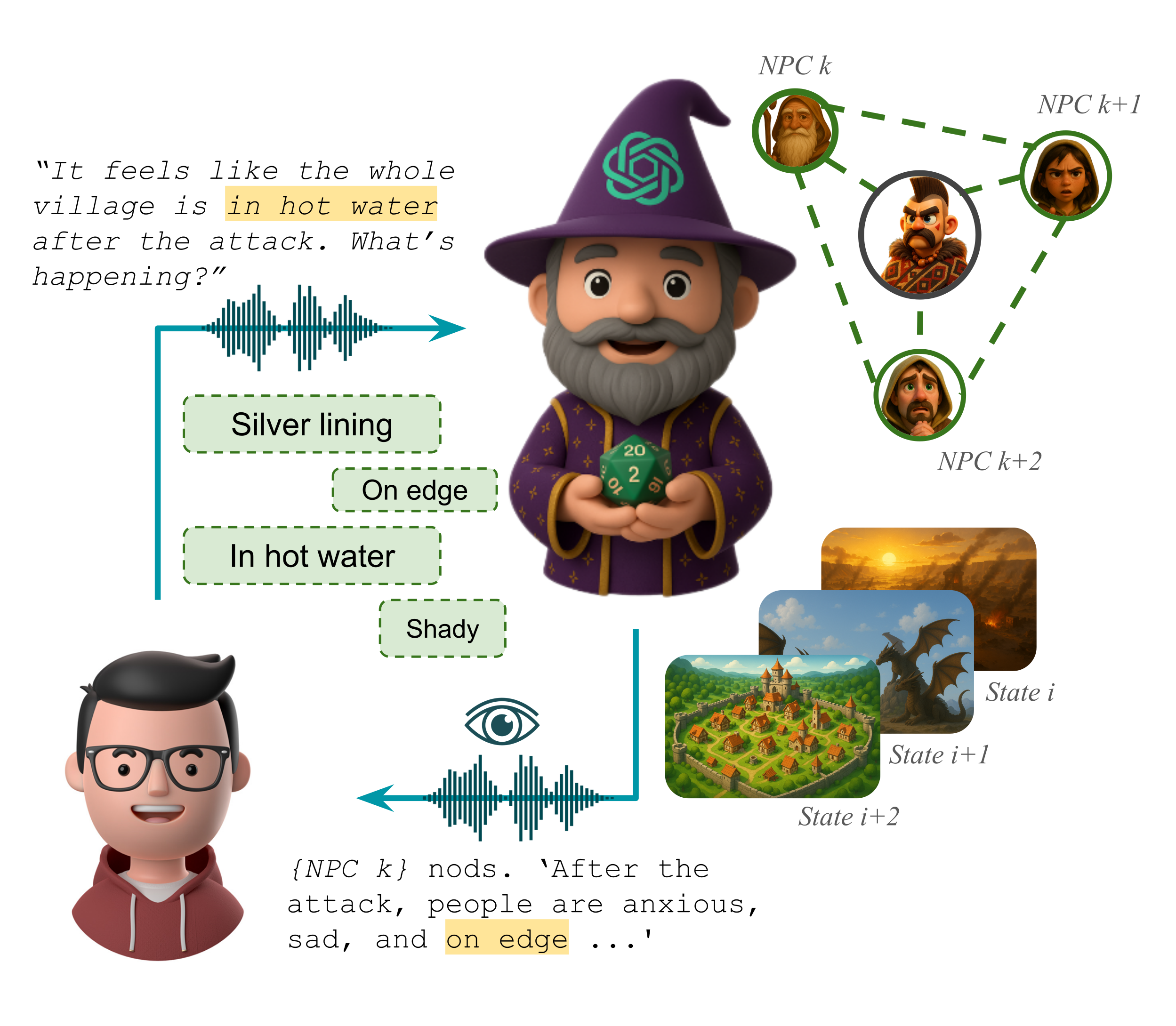}
    \caption{
 Overview of the LLM-powered RPG interaction loop. Spoken user input is transcribed by Whisper and sent to the GPT-4o Game Master along with the current game state, predefined scene assets as well as descriptions of NPCs. The Game Master dynamically advances the narrative, selects and displays the appropriate background and speaking/companion NPC(s) based on the player's actions and location, updates visual elements, and naturally recasts the learner’s target slang phrases in meaningful contexts when situationally appropriate.
    }
    \label{fig:system-architecture}
\end{figure}

Despite general proficiency as measured by standardized language assessments such as TOEFL~\cite{ets2024toefl} and IELTS~\cite{ielts2024scoring} many learners continue to face challenges in using language actively across diverse and spontaneous interaction settings. The speaking components of these exams tend to reflect structured, academic scenarios, offering only a partial view of a learner’s communicative ability in informal, imaginative, or less predictable conversations~\cite{ironsi2023vr}. As a result, students may score well on these exams yet continue to struggle with real-time conversations and natural syntax use in academic or social settings.

Language teaching methods have continuously evolved over the centuries. Traditional approaches began with the Grammar-Translation Method (GTM)~\cite{chang2011contrastive}, which emphasized grammatical structures and the translation of texts from the learner's native language into the target language. This was followed by the Audio-Lingual Method~\cite{mart2013audio}, which focused on memorization of dialogues and the use of repetitive drills to develop listening and speaking skills, with special emphasis on pronunciation. The focus then shifted to the Direct Method~\cite{dakhalan2024direct}, which emphasized instruction entirely in the target language. In this approach, comprehension was developed without explicit grammar instruction. An offshoot of the Direct Method was Content-Based Instruction (CBI), including models like CLIL (Content and Language Integrated Learning)~\cite{brinton1989content,dalton2011clil}, which borrowed from immersion programs and prioritized spoken communication through topic-based curricula rather than grammar or vocabulary drills. These traditional and early modern methods typically positioned the teacher as the central source of knowledge, with students playing passive roles. However, with the emergence of more modern pedagogical frameworks, the teacher’s role gradually shifted to that of a facilitator, supporting students in more active, collaborative learning such as cooperative learning~\cite{slavin1995cooperative,johnson1994learning}, which integrates academic and social activities, and Community Language Learning~\cite{curran1976counseling,richards2001approaches}, which emphasizes trust and student autonomy, embody this shift. The Task-Based Learning Method (TBLT)~\cite{ellis2003task,li2023review}, which has gained renewed attention in recent years, focuses on the use of authentic language to accomplish meaningful tasks in the target language. These modern methods have been further enhanced by technological advancements: virtual reality\cite{song2023optimizing}, artificial intelligence~\cite{Ruan2021AIchatbot}, and gamification~\cite{shen2024gamification} are increasingly being integrated into language education to create more immersive experiences that support engagement through learner-driven narratives. \\
Game-based learning, which has gained popularity in recent years, demonstrates moderate to large effects on cognitive, social, emotional, motivational, and engagement outcomes, making it a promising educational tool~\cite{lin2023game}. Studies have reported significant increases in both learner motivation and retention within gamified learning environments~\cite{kumar2024impact}, along with a notable reduction in language learning anxiety over time~\cite{Chen2022AnxietyGamified}. With the advancement of large language models (LLMs) \cite{openai2023gpt4}, including models like ChatGPT-4o, their application in language learning has become increasingly popular due to their capabilities in refining student writing and enhancing the quality of instructional feedback. Beyond instructional uses, ChatGPT serves as a self-learning tool, enabling independent language acquisition and cultural exploration \cite{ma2024chatgpt}. Concurrently, in game design, LLMs are naturally suited for facilitating natural language conversations, functioning as non-player character (NPC) dialogue systems, and assuming mastering or assistive roles within games \cite{gallotta2024llms}. This convergence presents a compelling opportunity to integrate these domains, leveraging LLMs to create immersive, interactive, and student-centric language learning experiences through game-based environments.\\ 
In this work, we aim to develop a task-based, interactive language learning system that leverages a large language model (LLM) agent as the master of an immersive, storytelling-driven \textit{Dungeons \& Dragons}-style role-playing game. The system is designed to encourage learners to engage in active spoken interaction using casual slang phrases as target expressions in their responses. Following the interaction, the system provides targeted quantitative and qualitative feedback on grammar and usage of the target vocabulary to support reinforcement and further language development.

\textbf{Our key contributions are as follows:}
\begin{itemize}
  \setlength\itemsep{0.7em}

  \item We designed and developed a domain-specific game-based language learning system that integrates an LLM-driven narrative with dynamic visual and verbal interaction cues. The system updates visual elements in real time based on story progression and learner performance, enabling an immersive experience.

  \item We incorporate a hybrid language learning approach that combines implicit and explicit feedback mechanisms. During gameplay, the agent subtly reinforces target expressions through natural recasting, and the system continuously tracks the learner’s usage of target slang words. The Post-interaction modules then provide explicit feedback through usage summaries, targeted corrections, and structured practice opportunities.
\end{itemize}

\section{Related Work}
In this section, we review the advancement of AI technologies in language learning, with a focus on the capabilities of large language models (LLMs), and explore the pedagogical potential of role-playing games as interactive, engaging environments that align well with communicative learning objectives.


\subsection{AI in Language Learning}
The history of AI in language learning largely began with chatbot-style systems, grounded in early advances in natural language processing (NLP) and machine learning~\cite{Caldarini2021ALS}. Early systems such as ELIZA~\cite{10.1145/365153.365168} were based on rule-based pattern matching and lacked the ability to interpret the meaning of user input. These systems simulated conversation but offered little in terms of contextual understanding or dynamic feedback. Subsequent developments included systems like Gengobot, a grammar dictionary chatbot designed for Japanese language learning~\cite{haristiani2019gengobot}, and CSIEC, which could generate simple communicative responses based on user input~\cite{jia2008csiec}. With the rise of deep learning and more advanced NLP techniques, conversational AI systems evolved to comprehend user intent, context, and sentiment, leading to more natural and flexible interactions. Commercial tools such as Quizlet, Ginger, and ProWritingAid began incorporating these capabilities to provide intelligent support in vocabulary acquisition, grammar correction, and writing enhancement~\cite{LLMbasedChatbots}.

Large Language Model (LLM)-based chatbots now offer a wide range of capabilities that can enhance the language learning process. Several studies have examined their potential to serve as real-time conversational partners, helping learners practice speaking and writing, understand idiomatic expressions and cultural references, and build fluency and confidence\cite{huang2022chatbots}.
In ~\cite{park2024align}, the authors developed an LLM-powered chatbot fine-tuned specifically for English as a Foreign Language (EFL) learners. The system initiates dialogue based on real-world themes (e.g., ordering food, job interviews). Learners speak or type responses, and the system generates contextually appropriate replies. It offers subtle feedback and rephrasing options, and learners can also ask grammar-related questions to receive in-depth explanations. In ~\cite{schmucker2024ruffle}, the authors introduced \textit{Ruffle\&Riley}, a conversational tutoring system that creates a dialogue-driven learning environment using LLMs. After selecting a lesson topic, the system automatically generates a tutoring script and facilitates a conversation between two AI agents, Ruffle (student) and Riley (tutor). Learners can observe or assume the role of Ruffle, interacting with Riley in a goal-focused dialogue. Throughout the session, the system prompts learners with reflective questions (e.g., “What do you think the main argument in this paragraph is?”), and provides feedback on engagement, comprehension, and suggested follow-ups.

In ~\cite{pan2024ellmat}, they present \textit{ELLMA-T}, an embodied LLM-powered avatar integrated into a social virtual reality (VR) environment. Learners enter immersive scenarios such as restaurants, airports, or classrooms, and engage in natural voice-based interactions. The agent is capable of real-time multi-turn conversations, adjusting its behavior based on learner proficiency and task context (e.g., giving a presentation versus casual dialogue). Sessions may be recorded and reviewed post-hoc, with feedback provided on vocabulary usage, missed speech opportunities, and overall performance. \\
In general, the recent trend in language education technology reflects a shift from isolated features such as grammar correction toward more immersive, interaction-driven experiences, an evolution that aligns closely with modern pedagogical approaches emphasizing communicative competence and learner-centered engagement.

\bibliographystyle{ACM-Reference-Format}

\begin{figure*}[t]
    \centering
    \includegraphics[width=0.8\textwidth]{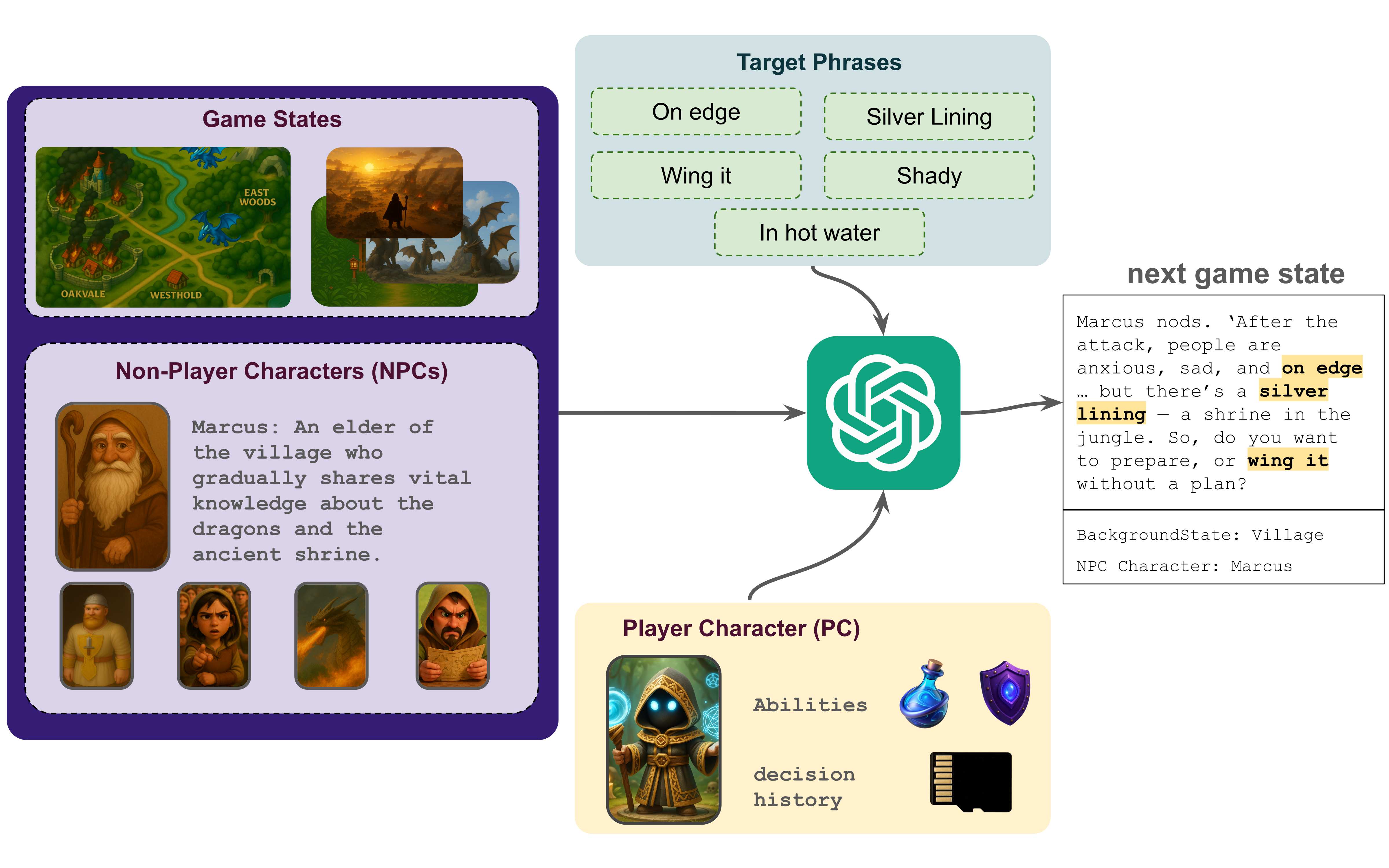} 
    \caption{Overview of Game Modules and LLM Narrative Generation: The agent receives the core game materials, including a set of NPC characters with brief descriptions to guide narrative progression, the current game state with possible locations and encounters, the set of target phrases, and the player’s chosen hero with its associated abilities and decision history. Based on these inputs, the model generates the next narrative segment, integrates the target phrases along with contextual explanations, and outputs the next game state and the next speaking NPC. The visual interface is then updated according to the agent’s interpretation.}
\end{figure*}

\subsection{Linguistic Assessment and Feedback Methods}
In the analysis of language proficiency, accuracy is considered one of the major dimensions ~\cite{inbook}. It refers to the extent to which a learner’s output adheres to the grammatical rules, vocabulary usage, and phonological or orthographic conventions of the target language. In this dimension, various types of feedback have been proposed to enhance learning outcomes. These include general feedback, which conveys overall information about errors (e.g., number, type); specific feedback, which provides sentence-based analysis with explanations and correction suggestions; and formative feedback, which summarizes errors and strengths while suggesting pathways for improvement~\cite{article}.
In terms of instructional approaches, language learning can be supported through both implicit and explicit strategies. Implicit feedback techniques, such as recasts, prompts, or clarification requests, encourage learners to use target phrases through natural interaction without direct correction, allowing patterns to be acquired subconsciously. In contrast, explicit feedback conveys corrective information directly, making the learner consciously attend to the target form. Studies have shown that these approaches differ in their effects: explicit feedback often leads to stronger immediate gains, whereas implicit feedback tends to yield better long-term retention\cite{LiShaofeng}.


\subsection{Role-Playing Games}
Role-playing games (RPGs) are a popular and widely varied category of games. Due to their broad and diverse formats, they can be difficult to define precisely; however, there are key properties that distinguish them from other game types. At their core, RPGs involve "playing a role", that is, embodying someone else’s perspective in an interactive process of defining and re-defining the state, properties, and contents of an imaginary game world~\cite{zagal2018definitions}. This emphasis on freedom, character embodiment, and player-centric mechanics, makes RPGs well aligned with modern language learning methodologies that prioritize communication, and contextual engagement. Several studies~\cite{ly2024role,khamouja2023importance,liu2010role,Yasin2024Impact} support this claim, showing that role-playing not only enhances language instruction, particularly in developing speaking skills, but also fosters positive learner behaviors, attitudes, and overall engagement with the learning process beyond just English proficiency.
\section{Hypotheses and Research Questions}
We hypothesize that integrating a task-based language learning framework within a role-playing game environment, guided by a large language model (LLM), can enhance learners’ active language use, engagement, and awareness of their own performance. Specifically, we propose that:

\begin{itemize}
    \item[\textbf{(H1)}] Learners who interact with the LLM-driven role-playing game will demonstrate a better understanding of the meanings of target words, higher accuracy in their use and associated syntactic structures.

    \item[\textbf{(H2)}] The immersive, narrative-driven structure of the game will result in higher learner engagement, as evidenced by self-reported measures and interaction activity data.
\end{itemize}

\section{Methodology}
In this section, we first provide a high-level overview of the user-facing design and interaction flow for both the experimental (RPG-based) and control (traditional AI classroom) conditions. We then describe the shared technical architecture that powers the system, followed by the experimental procedure, participant details, and evaluation measures.
\subsection{System Overview}

\subsubsection{Experimental Group}
The system operates as a dyadic~\cite{zhang2021dyadic}, spoken turn-based interaction, where communication is understood as a mutual and evolving exchange between two agents: the user and a GPT-based Master of the game. The main goal of this exchange is to collaboratively move the narrative forward. The Game Master (GM) establishes and dynamically adjusts the game world and its consequences, while the player drives the plot forward by making decisions and taking actions within that environment. The introduction of the game follows two principles: world-building and role embodiment. World-building is delivered through a short animated video that conveys the background and central mission. Role embodiment is fostered by letting participants choose one of four distinct heroes, each introduced with unique visuals and a short description of their special abilities. In each turn, the user receives the current narrative state of the game through the Master’s Dialogue Box, then responds verbally by pressing record button (Fig. 4). The system processes the input and advances the narrative, updating both verbal and non-verbal cues, such as the background environment and the NPCs’ persona, according to the evolving game state. A dedicated Practice Box (right side of the interface, Figure 4) displays the five selected target slang phrases along with their meanings and example sentences. The system automatically detects and tracks each phrase’s usage in the learner’s spoken input, updating a real-time color-coded progress indicator: red after one detected use and green after two. Our design goal is for each target phrase to be practiced at least twice during the 12-turn game; although usage does not affect narrative progression, the box serves as a persistent visual scaffold, and occasional pop-up reminders gently encourage learners to incorporate unused or once-used phrases. In addition to the explicit support from the Practice Box, the Game Master regularly uses the target slang phrases in its own dialogue throughout all game phases. These phrases appear in meaningful context or are paired with synonyms (e.g., “Do you really want to go in without preparation and wing it?”), giving learners repeated, authentic exposure to correct meaning and situational usage. An overview of the core modules and information flow that enable the Game Master to generate coherent, context-aware narrative responses is shown in Figure 2.\\

\begin{figure}[ht]
    \centering
    \includegraphics[width=1\linewidth]{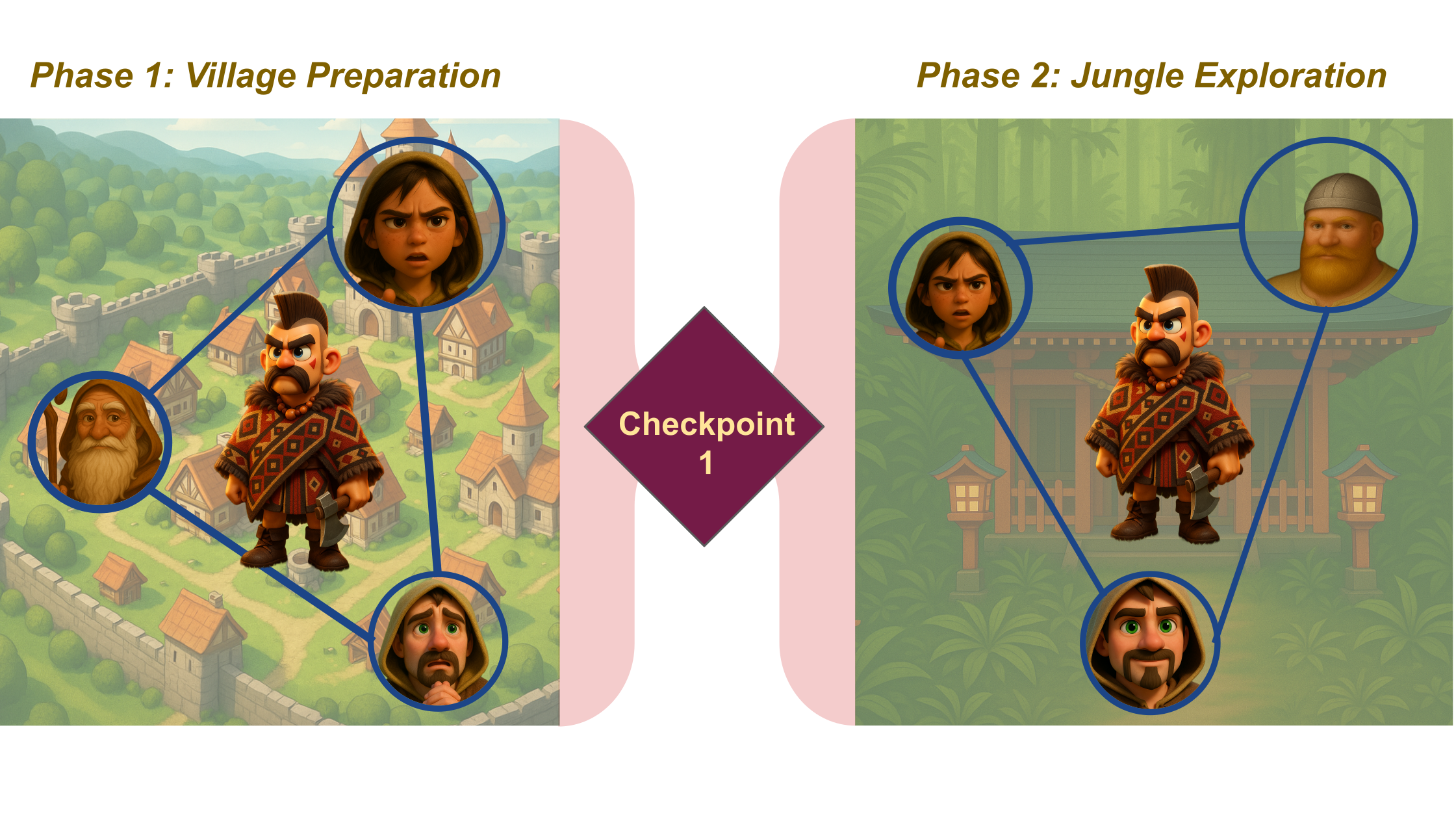} 
\caption{
Narrative flow structure of the role-playing game. The game is organized into sequential phases, each containing distinct communicative goals and challenges. Within each phase, player responses drive branching interactions (illustrated by multiple paths and NPC portraits). All paths converge at a checkpoint before advancing to the next phase.
}

    \label{fig:flow}
\end{figure}

The narrative structure follows a phase-based design, where each phase presents distinct goals and challenges.
The game's educational structure is built around three main phases. Throughout the game, the user interacts with a predefined set of in-game characters. as shown in Fig. ~\ref{fig:flow}:
\begin{itemize}
    \item \textbf{Phase 1: Preparation and Information Gathering} The initial stage requires players to engage in dialogue with characters to ask for guidance. This phase also allows players to form a group with the non playable characters (NPCs) for the subsequent adventure.
    \item \textbf{Phase 2: Exploration and Problem-Solving} This stage presents players with a series of puzzles and challenges. Successfully solving them grants specific abilities or skills for later use. Players can also explore the environment to find equipment and weapons.
    \item \textbf{Phase 3: Strategy and Leadership} The final phase involves a tactical battle. Victory depends on the player's ability to strategically use their character's acquired skills and resources to lead their party.
\end{itemize}

This three-phase design is intended to place language learners in diverse communication scenarios. The narrative structure allows for branching: while player decisions create different paths within each phase, all paths converge at a single point before the narrative advances to the next stage.
Final outcomes are determined based on a combination of checkpoint values and the Master’s assessment. 

\begin{figure}[ht]
    \centering
    \includegraphics[width=1.1\linewidth]{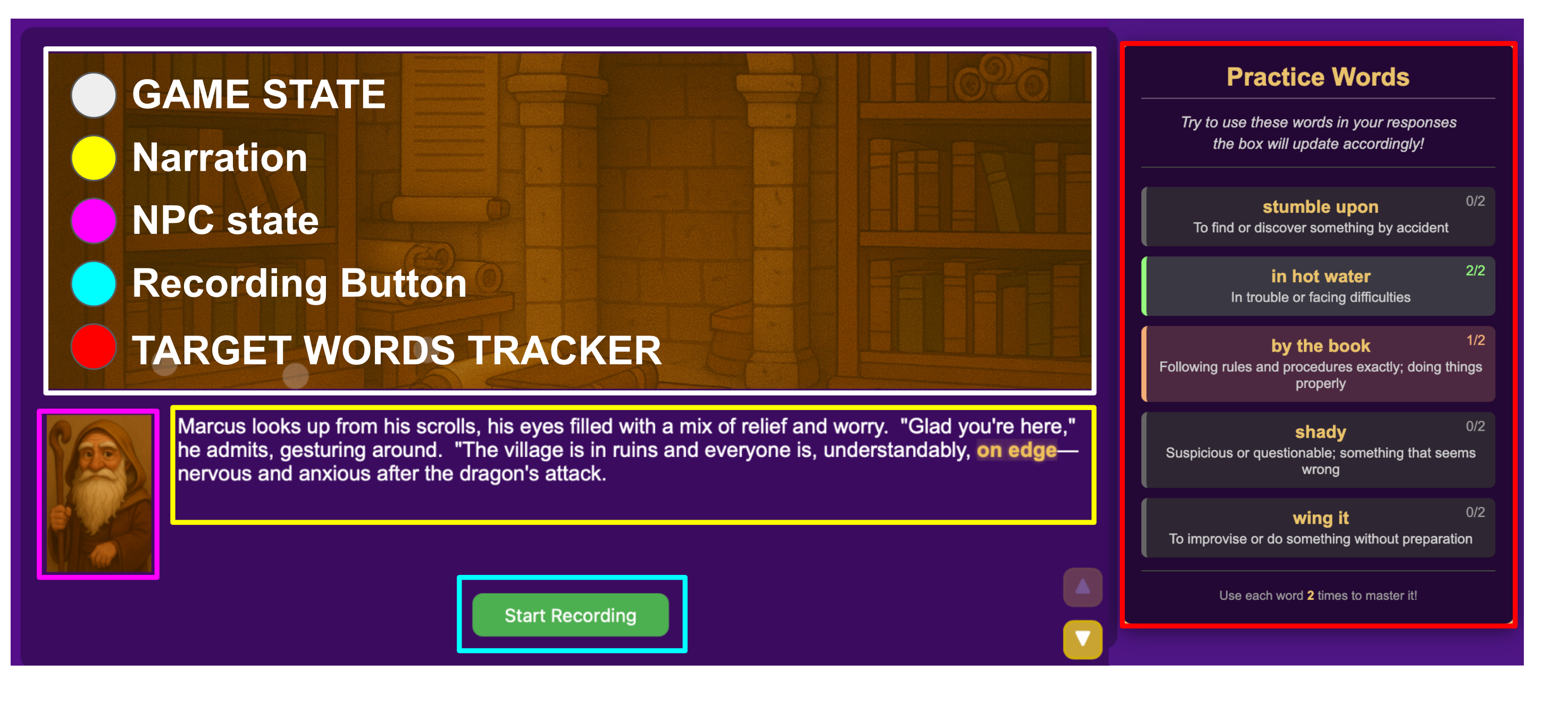} 
\caption{
Interface of the Game.
Colored circles highlight the main elements: 
\textbf{White} – Game State (top background), 
\textbf{\textcolor{yellow}{yellow}} - Narration / Mentor's Dialogue Box (bottom-middle, AI-generated text and prompts), 
\textbf{\textcolor{magenta}{magenta}} - NPC/mentor portrait (bottom-left, updates with tone), 
\textbf{\textcolor{cyan}{cyan}} - Recording Button, 
\textbf{\textcolor{red}{red}} - Target Words Tracker (right panel, lists the five slang phrases with meanings; each phrase turns red after one spoken use and green after two). 
}
    \label{fig:game-ui}
\end{figure}

\subsubsection{Control Group}
\begin{figure}[ht]
    \centering
    \includegraphics[width=\linewidth]{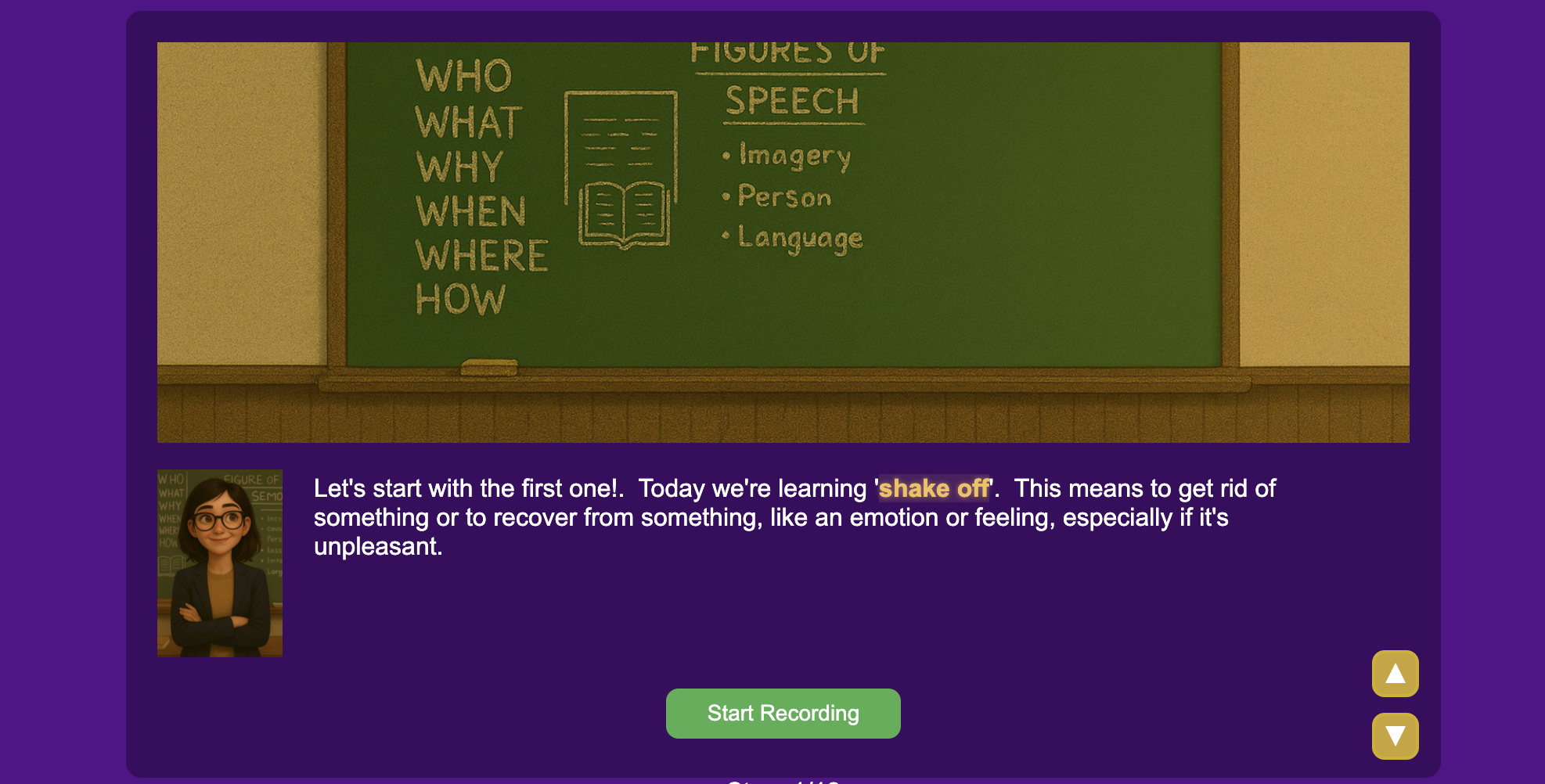} 
    \caption{
    The visual interface of the AI english class. 
    The \textbf{Mentor's Dialogue Box} (bottom-middle) displays the AI-generated narrative and prompts in response to user input.
    Players use the \textbf{Recording Button} (bottom-center) to speak their replies.
    }
    \label{fig:control-ui}
\end{figure}
The control group also follows a 12-turn, level-based structure divided into an introduction, ten steps of practicing target words, and a final summary step. In the introduction, the agent presents itself as an AI teacher, and during the ten practice steps, each target word is covered over two turns: first, the agent introduces the word and its meaning and asks the user to produce a sentence with it; then, after receiving the user’s response, the agent analyzes the sentence and provides explicit feedback on meaning accuracy or grammatical issues. It then proceeds to the next word. The session concludes with an outro summarizing the interaction.

\subsection{System Architecture}
To enable seamless, interactive gameplay and language learning, user speech is first transcribed using Whisper, a multilingual automatic speech recognition (ASR) model developed by OpenAI~\cite{radford2022whisper}. The transcribed input is then processed by GPT-4o, which generates narrative responses grounded in the current game context. These responses are vocalized using Google Cloud Text-to-Speech, producing real-time, natural-sounding audio output~\cite{googleTTS}. To support accessibility and reinforce comprehension, subtitles are displayed alongside audio responses, offering users dual verbal cues.

For non-verbal feedback, we utilize two main channels: dynamic game state visuals and character state indicators. Each game state (e.g., combat, negotiation, exploration) is associated with pregenerated visual assets, which are selected and updated based on the agent’s internal narrative evaluation. This visual system ensures that the environment remains responsive and contextually coherent throughout the interaction. 


\begin{figure}[ht]
    \centering
    \includegraphics[width=\linewidth]{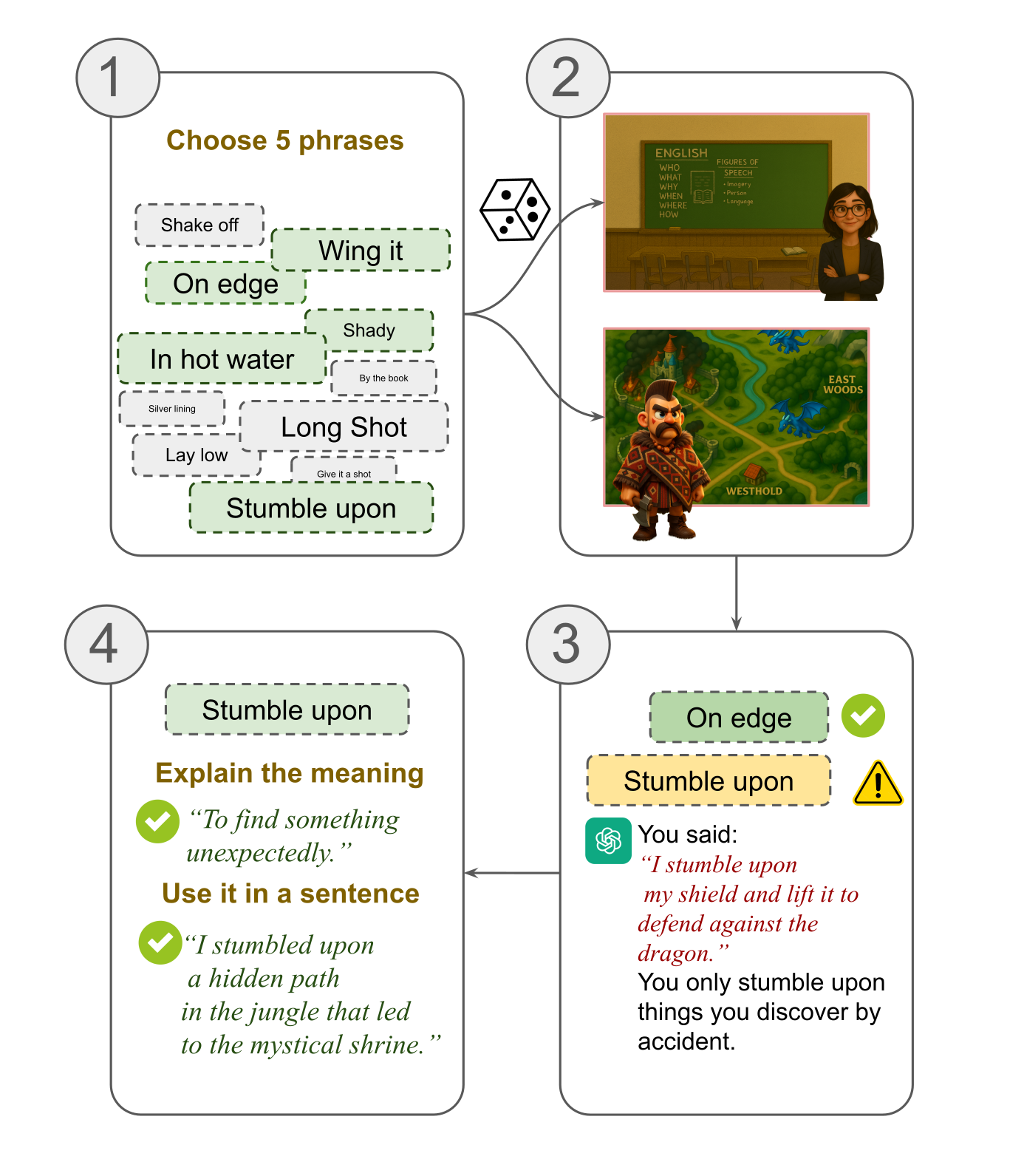} 
    \caption{
    Task Flow: (1) Initial assessment: participants are shown 12 slang phrases and select 5 they are unfamiliar with or less comfortable using. (2) Participants are then randomly assigned to either the experimental or control group and complete a 12-turn activity with the AI agent. (3) After the activity, explicit post-interaction feedback is provided based on the participant’s responses. (4) Final assessment: participants define and use the phrases again.
    }
    \label{fig:taskflow}
\end{figure}
 
\subsection{Experimental Setup}
\subsubsection{Participants}
We recruited 16 international students (aged 21-30), all of whom had met the English language proficiency requirement for university admission (minimum TOEFL score of 90) and were actively engaged in English learning. Participants were randomly assigned to either the experimental group (LLM-assisted RPG interaction) or the control group. Two participants failed to complete the full session (one from each condition) and were excluded from the analysis. The remaining 14 participants were randomly assigned to either the experimental condition (RPG interaction, n=7) or the control condition (AI classroom, n=7).
\subsubsection{Task Flow}
The overall task flow is the same for both the experimental and control groups and is shown in Figure 6. First, the system presents 12 predefined slang phrases and asks participants to select 5 phrases they are unfamiliar with or less comfortable using. These phrases were chosen based on recommendations from English instructors at Purdue who work with international students and frequently observe difficulties with these expressions. After selecting 5 phrases, the system asks participants to indicate whether they are completely unfamiliar with each word, somewhat familiar with it, or able to guess its meaning. To more precisely assess partial knowledge, participants who claimed they partially knew a word were asked to verbally define its meaning and use it in a sentence.\\
After the initial assessment, participants were assigned to either the control or experimental group to practice their selected phrases with the AI agent, either through a virtual English class or through the game setup described in the previous sections. Both groups completed a 12-turn dialogue with the AI agent, which lasted approximately 30 minutes on average.\\
The activity is followed by a post-interaction feedback phase, whose goal is to provide a general assessment of the interaction with a focus on the Accuracy dimension. The agent is prompted to deliver multi-level written feedback, which includes: (1) general feedback summarizing the number and types of errors; (2) specific feedback offering sentence-level corrections and explanations; and (3) formative feedback that revisits the target word list, evaluates whether each term was used, and, when applicable, analyzes its correctness and appropriateness while optionally providing examples or revised uses. Then in a immediate assessment, participants are asked to define the meaning of each phrase and use it in a sentence, and they receive brief textual feedback on their usage. To assess retention, a delayed post-test was conducted one week later, where participants were again asked to define and use the target phrases in sentences.

\subsubsection{Measures} 
In this study, we use multiple sources for analyzing linguistic performance and engagement. For the language-related measures, we evaluate participants’ understanding and use of the target words across the initial assessment and the immediate post-test. These measures include word meaning accuracy and contextual appropriateness, meaning that even if there are some issues such as grammar, the intended meaning and situational use are correct. Engagement was evaluated using a self-reported survey administered immediately after the post-test, supplemented by thematic analysis of participants’ open-ended responses.

\section{Results \& Discussion}
In this section, we present both qualitative and quantitative findings. We organize our preliminary results into two parts: (1) language learning performance, with immediate post-intervention gains (2) engagement outcomes based on self-reported surveys; and qualitative feedback and observations. 
\subsection{Language Learning Performance}
To evaluate linguistic performance, we quantified participants' understanding of the target phrases using a three-tiered scoring rubric. A score of 1.0 was awarded for a correct and comprehensive definition, 0.5 for a partially correct definition, and 0.0 for an incorrect definition or if the participant was unable to provide one. For sentence accuracy, a score of 1.0 was assigned when both syntactic correctness and contextual appropriateness were satisfied; a score of 0.5 was given when the intended meaning and context were appropriate but syntactic errors were present; and a score of 0.0 was assigned when the sentence failed to convey appropriate meaning or context. The assessment of semantic accuracy for both the definitions and the participant-generated sentences was performed using GPT-4o. This evaluation was calibrated to prioritize contextual appropriateness and intended meaning; grammatical errors that did not impede comprehension did not result in a zero score but were reflected as partial credit in the rubric.

Analysis of the pre-test data indicated a baseline imbalance between cohorts; the experimental group demonstrated a slightly higher initial proficiency with the target phrases compared to the control group, scoring 1.50 compared to 1.12. To account for this discrepancy and accurately measure gains relative to each group's starting point, we calculated a normalized learning rate. This approach mitigates the ceiling effect for participants with higher pre-test scores. The formula is defined as:

\begin{equation}
   \text{Learning Rate}  = \frac{\text{PostScore} - \text{PreScore}}{\text{MaxScore} - \text{PreScore}}
\end{equation}

where scores ranged from 0 (completely incorrect or no response) to 5 (perfect definition and appropriate contextual use), making 5 the maximum possible score. We applied this formula to the average scores for both definition accuracy and correct sentence generation. The resulting normalized learning rates for both cohorts are presented in Table~\ref{tab:learning-rate}. The data indicates that the RPG group achieved a higher learning rate in both measures. Sentence accuracy is slightly higher than definition accuracy because participants were often able to infer appropriate contextual usage during interaction, even when they struggled to recall precise definitions.

\begin{table}[htbp]
\centering
\caption{Normalized learning rate (immediate post-test) across groups and assessment tasks. Both groups show comparable improvements, with the RPG condition achieving slightly higher gains.}
\label{tab:learning-rate}
\small
\setlength{\tabcolsep}{5pt}
\renewcommand{\arraystretch}{1.25}
\begin{tabular}{p{2.8cm}cc}
\noalign{\hrule height 0.8pt}
\textbf{Assessment Task} & \textbf{Control} & \textbf{RPG} \\
\noalign{\hrule height 0.8pt}
\rowcolor{red!8}
Definition Accuracy & 0.82 & \textbf{0.88} \\
\rowcolor{red!12}
Sentence Accuracy & 0.92 & \textbf{0.95} \\
\noalign{\hrule height 0.8pt}

\end{tabular}
\end{table}

To evaluate the long-term effectiveness of the learning intervention, vocabulary retention is assessed one week after the immediate post-test. The retention rate quantifies the proportion of newly acquired knowledge that the learner successfully maintained over this period, and is calculated using the following formula:
\begin{equation}
    \text{Retention Rate} = \frac{\text{PostScore}_{1\text{ week}} - \text{PreScore}}{\text{PostScore} - \text{PreScore}}
\end{equation}
A baseline retention rate of 1 means all immediate gain is retained, while 0 means performance has returned to the pre-test level. Because the formula tracks relative change, a score greater than 1 represents additional proficiency gained after the immediate post-test, whereas a negative score would indicate a regression below the participant's original baseline. The retention rates for both groups are also shown in Table~\ref{tab:retention-rate}.

\begin{table}[htbp]
\centering
\caption{Retention rate (one-week delayed post-test). The RPG condition shows 21\%–27\% higher retention than the control group, whereas the control group exhibits decline.}
\label{tab:retention-rate}
\small
\setlength{\tabcolsep}{5pt}
\renewcommand{\arraystretch}{1.25}
\begin{tabular}{p{2.8cm}cc}
\noalign{\hrule height 0.8pt}
\textbf{Assessment Task} & \textbf{Control} & \textbf{RPG} \\
\noalign{\hrule height 0.8pt}
\rowcolor{blue!8}
Definition Accuracy & 0.86 & \textbf{1.04} \\
\rowcolor{blue!12}
Sentence Accuracy & 0.83 & \textbf{1.05} \\
\noalign{\hrule height 0.8pt}
\end{tabular}
\end{table}

While both cohorts demonstrated substantial improvement, these findings suggest that the immersive, task-based RPG framework was comparatively more effective in enhancing learners' semantic understanding and contextual application of the target phrases. Consequently, the retention results demonstrate that the RPG framework effectively combats typical memory decay. While participants in the control group forgot a portion of what they learned over the following week, the RPG participants retained their immediate learning in both defining and using the target vocabulary. This suggests that the engaging nature of the game makes the vocabulary more memorable, leading to durable, long-term comprehension.


    In the post-test sentence-creation task, the two most highly engaged participants from RPG group (identified by response length and narrative creativity) spontaneously transferred elements of the game’s fictional world into their examples, incorporating references to the game. In contrast, the majority of participants across both conditions contextualized the target slang within familiar daily routines, For example P6 said: \\ 
\begin{quote}

\textit{"As I'm reaching the deadline, I think there's no time and I need to wing it."}
\end{quote}
or P8: 
\begin{quote}
\textit{"Every day after work, I take a shower. It helps me shake off the stress."}
\end{quote}

This pattern suggests that, while deep story immersion can foster creative transfer of the game context for highly engaged learners, most users default to real-life scenarios. A promising direction for future work would therefore be personalized storytelling that adapts the narrative to each learner’s own daily routines and concerns, which may yield even stronger and more personally relevant learning outcomes.

\subsection{Engagement Outcomes}
To evaluate subjective engagement and user experience, we analyzed participant responses to a post-activity survey. All self-reported data was quantified using a five-point Likert scale, corresponding to the following mapping: (1) "Strongly disagree," (2) "Somewhat disagree," (3) "Neither agree nor disagree," (4) "Somewhat agree," and (5) "Strongly agree."

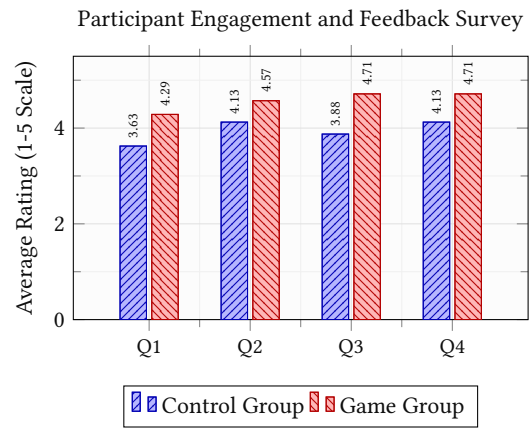
\begin{figure}[h!]
\centering
\begin{tikzpicture}
\begin{axis}[
    title={Participant Engagement and Feedback Survey},
    width=0.9\columnwidth,
    height=0.6\columnwidth,
    ybar,
    bar width=10pt,
    ymin=0, ymax=5.5,
    ylabel={Average Rating (1-5 Scale)},
    nodes near coords,
    enlarge x limits=0.25,
    legend style={at={(0.5,-0.25)}, anchor=north, legend columns=-1},
    symbolic x coords={Q1,Q2,Q3,Q4},
    xtick=data,
    xticklabels={
        Q1,
        Q2,
        Q3,
        Q4
    },
    xticklabel style={text width=3.5cm, align=center, font=\small},
    nodes near coords align={vertical},
    nodes near coords style={font=\tiny, rotate=90, anchor=west}, 
    fill opacity=0.95,                            
    grid=both,                                 
    minor tick num=1,                          
    grid style={line width=0.2pt, draw=black!12},   
    minor grid style={line width=0.15pt, draw=black!6}, 
    axis background/.style={fill=gray!2},      
]

\addplot[
    ybar,
    bar width=10pt,
    fill=blue!30,                  
    draw=blue!70!black,            
    postaction={
        pattern=north east lines,  
        pattern color=blue!70!black
    }
] coordinates {
    (Q1,3.625)
    (Q2,4.125)
    (Q3,3.875)
    (Q4,4.125)
};

\addplot[
    ybar,
    bar width=10pt,
    fill=red!30,
    draw=red!70!black,
    postaction={
        pattern=north west lines,
        pattern color=red!70!black
    }
] coordinates {
    (Q1,4.285714)
    (Q2,4.571428)
    (Q3,4.714285)
    (Q4,4.714285)
};

\legend{Control Group, Game Group}
\end{axis}
\end{tikzpicture}
\caption{Comparison of self-reported engagement and feedback between the Control and Game groups. 
The four measured items correspond to: 
(Q1) ``I was engaged during the activity,'' 
(Q2) ``The examples and contexts helped me understand the vocabulary,'' 
(Q3) ``I had enough opportunity to use the vocabulary,'' and 
(Q4) ``The way the AI mentor/Game Master used the target words helped me understand them.''}

\label{fig:engagement-chart}
\end{figure}

As illustrated in Fig.~\ref{fig:engagement-chart}, the experimental (RPG) group reported a more positive experience than the control group across all measures. Specifically, participants in the game group perceived the environment as more engaging, found the narrative contexts more useful for understanding vocabulary, and felt the system provided a superior opportunity to practice the target phrases.

Beyond the quantitative ratings, qualitative feedback from the experimental group further elucidates why the game was perceived as more engaging and effective. A recurring theme was the system's narrative autonomy and its impact on contextual application. Participants appreciated that the game's adaptive storyline provided a meaningful framework for deciding how and when to use the target phrases. As P1 explained:
\begin{quote}
    \textit{"It adapts to the user's response and let's the user create his own unique story (though within the general synopsis of the story), which I find very helpful in letting me decide where to use what words."}
\end{quote}

A second prominent theme was the value of active language production over the passive reception of information. Participants felt the game's task-based mechanics, which required them to use the vocabulary to advance, were more effective than traditional drill-based methods. P7 articulated this distinction:
\begin{quote}
    \textit{"It didn’t feel like the vocabs were spoon fed to me since I was required to produce language through making sentences and speaking them out."}
\end{quote}

These comments suggest that the RPG framework successfully created an environment where learners were intrinsically motivated to use the target language.

\section{Conclusion}

In this work, we designed, developed, and evaluated an LLM-powered role-playing game that employs a task-based language learning framework to enhance learners’ acquisition and spontaneous use of casual English slang vocabulary.
Our findings demonstrate that while both interactive methods, role-playing style and traditional virtual classroom instruction, are effective, the task-based RPG framework yields superior results in both linguistic performance and learner engagement. Quantitatively, the experimental RPG cohort exhibited higher normalized growth rates for both definition accuracy, and correct sentence generation. Furthermore, the RPG group demonstrated significantly higher active vocabulary usage, applying the target phrases much more frequently during the activity than the control group.
The qualitative survey results suggest that participants perceived the game environment as more engaging. Furthermore, the data indicates that the examples and narration provided within the game were more effective in supporting their learning compared to the control group.

\section{Limitations \& Future Work}

Having more participants would yield more robust and reliable statistical insights. For future work, several directions are possible: expanding feedback modalities to include not only accuracy but also complexity, fluency, and pronunciation of the target phrases; and exploring alternative storytelling and game-mechanic designs that can be implemented and tested.

\bibliography{references}
\end{document}